\documentclass[%
 aip,
 amsmath,amssymb,
 reprint,%
]{revtex4-1}

\usepackage{graphicx}
\usepackage{dcolumn}
\usepackage{bm}

\usepackage[utf8]{inputenc}
\usepackage[T1]{fontenc}
\usepackage{mathptmx}
\usepackage{array}
\usepackage{longtable}
\usepackage{etoolbox}
\usepackage{booktabs}
\usepackage[flushleft]{threeparttable}
\usepackage{siunitx, upgreek}
\usepackage[version=4,arrows=pgf-filled,
mathfontname=mathsf]{mhchem}

\DeclareSIUnit\bohr{\text{\ensuremath{a_0}}}      
\DeclareSIUnit\autime{\text{\ensuremath{t_0}}}   
\DeclareSIUnit\hartree{\text{\ensuremath{E_\mathrm{h}}}} 

\makeatletter
\def\@email#1#2{%
 \endgroup
 \patchcmd{\titleblock@produce}
  {\frontmatter@RRAPformat}
  {\frontmatter@RRAPformat{\produce@RRAP{*#1\href{mailto:#2}{#2}}}\frontmatter@RRAPformat}
  {}{}
}%
\makeatother
\begin{document}

\preprint{AIP/123-QED}

\title[Ab initio Simulations of \ce{EMI-BF4} Neutral-Surface Interactions in Electrospray Thrusters]{Ab initio Simulations of \ce{EMI-BF4} Neutral-Surface Interactions in Electrospray Thrusters}
\author{N. Laws}
 \email{nrl49@cornell.edu}
\author{E. Petro}
\affiliation{%
Sibley School of Mechanical \& Aerospace Engineering, Cornell University, Ithaca, New York
}%

\date{\today}

\begin{abstract}

Electrospray thrusters promise compact, high specific impulse propulsion for small spacecraft, yet ground characterization remains confounded by secondary species emission and incomplete diagnostics of neutral products. To address these limitations, we perform energy-resolved mixed quantum/classical (QM/MM) \textit{ab initio} molecular dynamics (MD) of neutral 1-ethyl-3-methylimidazolium tetrafluoroborate, \ce{EMI-BF4}, colliding with \ce{Au} extractor surfaces with impact energies from $10$ to $100$ \unit{\eV} to resolve fragment species spectra, charge states, kinetic energy partitioning, and scattering geometry. The simulations reveal a three-stage sequence with impact energy: the low energy regime, $10$ to $20$ \unit{\eV}, which favors ionic dissociation, intermediate energy regime, between $30$ to $40$ \unit{\eV}, opens a neutralization window, and high energy regime, greater than $50$ \unit{\eV}, drives covalent fragmentation into many light products with mixed charge states. Fractional energy distributions show a transition from few-body, energy-concentrated outcomes in the low energy regime to many-body, energy-dispersed outcomes in the high energy regime. Deflection angle distributions exhibit a strong mass-to-angle anti-correlation such that heavier fragments favor small deflection, whereas lighter fragments populate larger deflection angles. The fraction of transient metastables peaks near $50$ \unit{\eV}, coinciding with abundant neutral fragment production. Importantly, neutral bombardment still produces charged secondaries at the target even when the upstream ion plume is fully suppressed by a decelerating electrode. These findings provide a basis for de-biasing facility measurements by pairing tandem time-of-flight secondary ion mass spectrometry and residual gas analyzer with suppression-bias corrections to inform the design of electrospray thrusters that reduce interception and contamination on extractor surfaces.

\end{abstract}

\maketitle

\section*{Nomenclature}
\vspace{-.5em}
{
\renewcommand{\arraystretch}{1.2}%

\begin{longtable}{rl{c}}

$n$ & = Cluster solvation number \\
$\Delta t$ & = Maximum microcanonical ensemble timestep \\
$KE_i$ & = Prescribed incident projectile kinetic energy \\
$v_{CoM}$ & = Center-of-mass speed \\
$M$ & = Molar mass \\
$m$ & = Incident projectile mass \\
$N_A$ & = Avogadro's constant \\
$V_0$ & = Initial orientation vector \\
$V_i$ & = Target orientation vector \\
$u$ & = Rotation axis direction \\
$k_d$ & = Unit rotation axis direction \\
$\theta$ & = Rotation angle between $V_0$ and $V_i$ \\
$R$ & = Rotation matrix about $k$ \\
$K$ & = Skew-symmetric cross-product matrix of $k$ \\
$\delta E_{max}$ & = Maximum relative deviation of the total energy \\
$E_{tot,i}$ & = Instantaneous total energy at MD step $i$ \\
$\left<E_{tot}\right>$ & = Time average of the total energy \\
$\phi_x, \phi_y, \phi_z$ & = Projectile orientation Euler angles\\
$\chi_\mu \left(r_e\right)$ & = Contracted Gaussian-type basis functions \\
$\mu$ & = Atomic center \\
$r_e$ & = Electron position \\
$k_p$ & = Primitive Gaussian function index \\
$d_{\mu k}$ & = Contraction coefficients \\
$\alpha_{k}$ & = Primitive exponents \\
$R_{\mu}$ & = Nuclear position \\
$E_{XC}$ & = Exchange correlation energy \\
$\omega$ & = Range-separation parameter \\
$E_{X}^{HF,LR}\left(\omega\right)$ & = Long-range Hartree-Fock exchange \\
$E_{X}^{GGA,LR}\left(\omega\right)$ & = Long-range semilocal GGA exchange \\
$E_{X}^{GGA,SR}\left(\omega\right)$ & = Short-range semilocal GGA exchange \\
$E_{C}^{GGA,LR}\left(\omega\right)$ & = Long-range semilocal GGA correlation \\
$E_{C}^{GGA,SR}\left(\omega\right)$ & = Short-range semilocal GGA correlation \\
$E_{C}^{GGA}$ & = Semilocal GGA correlation \\
$E_{C}^{NL}$ & = Nonlocal correlation \\
$a$ & = Exact-exchange mixing parameter \\
$U_{ij}$ & = Exact-exchange mixing parameter \\
$r_{LJ}$ & = Interatomic separation \\
$\varepsilon$ & = Well depth \\
$\sigma$ & = Cutoff parameters \\
$k_0$ & = Initial NVE trajectory frame index \\
$k$ & = NVE trajectory frame index \\
$N$ & = Number of atoms in a neutral projectile \\
$r_i^{(k)}$& = Position vector of atom $i$ at frame $k$ \\
$d_{ij}^{(k)}$& = Interatomic distance matrix at frame $k$ \\
$d_b$ & = Distance threshold defining bond adjacency \\
$A_{ij}^{(k)}$& = Bond adjacency matrix \\
$G^{(k)}$& = Undirected graph at frame $k$ \\
$C^{(k)}$& = Number of connected atoms in $G^{(k)}$ \\
$k_{frag}$ & = NVE trajectory fragmentation frame index \\
$q_{A}$ & = Fitted partial charge on atom $A$ \\
$w_{i}$ & = Weight for grid point $i$ \\
$V^{QM}_{i}$ & = Electrostatic potential at grid point $i$ \\
$r_{A_i}$ & = Distance from atom $A$ to grid point $i$\\
$\lambda_{A}$ & = Restraint coefficient applied to atom $A$\\
$q^0_{A}$ & = Target charge for atom $A$\\
$Q_{k}\left(t\right)$ & = Fragment net charge at time $t$\\
\end{longtable}
}

\section*{Acronyms}
\vspace{-.5em}
{
\renewcommand{\arraystretch}{1.2}%

\begin{longtable}{rl{c}}

IL           & = Ionic liquid \\
SSE           & = Secondary species emission \\
SEE           & = Secondary electron emission \\
SIE           & = Secondary ion emission \\
TOF           & = Time-of-flight \\
SIMS           & = Secondary ion mass spectrometry \\
QCM           & = Quartz crystal microbalance \\
RGA           & = Residual gas analyzer \\
MD           & = Molecular dynamics \\
LJ           & = Lennard-Jones \\
EAM           & = Embedded atom method \\
PIR           & = Pure ion regime \\
CEX           & = Charge exchange \\
QM/MM           & = Quantum mechanics/molecular mechanics \\
GAPW          & = Gaussian augmented plane wave \\
DFT          & = Density functional theory \\
QM & = Quantum mechanics \\
GGA & = Generalized gradient approximation \\
ADMM-Q & = Auxiliary density matrix method \\
MM & = Molecular mechanics \\
RESP & = Restrained electrostatic potential \\
RMSE & = Root mean square error \\
RRMSE & = Relative root mean square error \\
RPA & = Retarding potential analyzer \\
\end{longtable}
}

\section{\label{sec:introduction}Introduction}

Propulsion systems for miniaturized satellites are often limited by mass, efficiency and lifetime in long-term missions. Electrospray thrusters mitigate the constraints at this scale with compact hardware, high propellant efficiency, and no moving parts. In these devices, capillary action feeds room-temperature ionic liquid (IL) propellants such as 1-ethyl-3-methylimidazolium tetrafluoroborate (EMI-BF4) \cite{lozano2005efficiency} to a series of porous emitters. A strong field between the emitters and a shared extractor electrode deforms the IL meniscus at each emitter into a Taylor cone. The field enables ion emission from the IL through the extractor aperture, yielding specific impulses between 100 to 10000 \unit{\s} depending on emission mode and operating point \cite{ziemer2009performance}.

\subsection{Propellant Overspray and Extractor Grid Lifetime}

Propellant overspray onto the extractor and accelerator grids is regarded as the primary lifetime and ground characterization limitation in electrospray thrusters \cite{collins2019assessment, thuppul2020lifetime}. Secondary species emission (SSE) leads to the accumulation of backspray material on mission-critical hardware that can distort field lines, increase beam divergence, and trigger electrical shorts that reduce performance and lifetime \cite{krejci2017emission}. Off-axis particles that collide with the extractor or vacuum chamber surfaces provoke SSE, including electrons, ions, ion clusters, and droplets.

Electrospray ion sources operated in the pure ion regime (PIR) predominantly emit singly charged ions with low solvation number $n < 3$. In this mode, a high electric field localized at the meniscus near the emitter apex drives ion evaporation from room temperature IL’s such as \ce{EMI-BF4}. By reversing the applied polarity, the dominant charged species produced is changed from beams of cations, such as \ce{EMI+}, in the positive mode or anions, such as \ce{BF4-}, in the negative mode without changing the underlying evaporation mechanism. Apex fields required for ion evaporation are on the order of $10^9$ ${V}/{m}$ \cite{lozano2005ionic}, consistent with electrically assisted ion evaporation from an IL meniscus \cite{gallud2022emission}. The emitted species are indexed by the solvation number $n$, which counts the number of neutral \ce{EMI-BF4} pairs bound to a central ion. Particles with $n = 0$ are monomers, $n = 1$ are dimers, $n = 2$ are trimers, and so forth for larger ion clusters. Because charge-to-mass ratio decreases with $n$ \cite{krejci2017emission, ma2021plume, petro2022multiscale, jia2022quantification}, the low $n$ ions and clusters set the baseline for propulsive performance and define the species most likely to intercept metallic surfaces. After emission, ions are accelerated roughly through the source potential between the emitter and extractor and then exit with a finite angular spread beyond the grid aperture. Multiscale plume models \cite{hampl2022comparison, whittaker2023modeling, smith2024propagating} and array experiments \cite{guerra2016spatial, krejci2017emission, natisin2019performance} show that the number and distribution of active emission sites influence beam divergence and the fraction of current intercepted by nearby hardware. As the plume propagates, intra-plume processes provoke fragmentation that populate the plume with neutrals spanning energies from thermal to nearly the full beam potential \cite{collins2019assessment, bendimerad2022molecular, geiger2024qcm, shaik2024characterization}. Therefore, linking plume physics to extractor grid lifetime requires understanding of what species are produced when IL constituents collide with metallic surfaces.

\subsection{Facility Effects and Secondary Species Diagnostics} \label{sec:surface_interactions}

Unlike in on-orbit operation, vacuum-chamber experimental characterization inherently intercepts the plume on beam targets, diagnostics, and chamber walls, making plume-surface interactions and SSE an unavoidable facility effect. Neutrals formed upstream of the target are not influenced by electrostatic biases, instead they retain the velocity of the parent ion cluster at the moment of breakup and can collide with nearby hardware and chamber surfaces at thermal velocity to full potential. SSE from surfaces impacted by electrospray plumes is measured with experimental diagnostics that separately access charged and neutral products. Uchizono et al. quantifies SSE as a first-order facility effect for electrospray thrusters and must be explicitly characterized during ground testing \cite{uchizono2021role}. Uchizono place the onset of secondary electron emission (SEE) near $0.44$ ${eV}/{u}$ and of secondary ion emission (SIE) near $1$ ${eV}/{u}$ for IL monomers and small clusters. Uchizono frames these metrics as motivation for the use of instrumentation that can resolve composition, polarity, and energy dependencies of the emitted secondary species.

Time-of-flight secondary ion mass spectrometry (TOF-SIMS) is an experimental instrument that has recently been applied to directly measure charged electrospray ion plume secondaries emitted from beam targets. The instrument directly resolves the mass-to-charge composition of charged secondary species produced when an ion-mode plume impinges on metallic targets representative of grid materials \cite{geiger2025secondary, hofheins2025electrospray}, including surfaces that can become conditioned by deposited propellant during operation \cite{hofheins2025iepc}. In these measurements, molecular ion clusters and co-streaming neutrals fragment upon collision with surfaces generating SSE observed in both polarities and include sputtered target ions as well as IL fragments such as \ce{B+}, \ce{H+}, \ce{CH2+}, \ce{F-}, \ce{HF-}, and \ce{BF4-} \cite{geiger2025secondary, hofheins2025electrospray, hofheins2025iepc}. However, TOF-SIMS is intrinsically limited to secondary ions and records positive-ion and negative-ion spectra in separate modes, indicating the need for complete polarity-resolved measurement of propellant-surface fragments.

To understand plume deposition, quartz crystal microbalances (QCM's) have been used. Geiger et al. exposed a $6$ \unit{\MHz} QCM to varying impact energy \ce{EMI-BF4} plumes to show that the transition between deposition and sputtering occurs with increasing energy \cite{geiger2024qcm}, which provides the impact energy threshold to near onset and \ce{Au} sputtering of $10$ to $100$ \unit{eV} targeted in this work. Furthermore, Geiger measured emission of charged secondaries from \ce{EMI-BF4} plumes where the ion plume was electrostatically suppressed so that only neutrals reach the target \cite{geiger2025secondary}. Geiger demonstrated that charged secondaries were produced when neutrals impacted the surface, which implies that neutral impact channels are an important consideration in SSE.  




Residual gas analyzer (RGA) complements these charged particle diagnostics by sampling neutrals in situ via electron impact ionization in the analyzer volume. During \ce{EMI-BF4} operation, Shaik et al. used an RGA to identify abundant neutral byproducts \cite{shaik2024characterization} such as \ce{HF}, \ce{H2}, and \ce{BF}, which provides constraints on volatile collision products that are invisible to TOF-SIMS and not separable by suppression-bias probes. The diagnostics discussed establish that, in the thermal to thousands of \unit{\eV} regime characteristic of IL neutrals, neutral incidence with metallic target collisions molecular secondaries.

\subsection{Computational Modeling of Electrospray Surface Collisions}

MD simulations have been used to model propellant-surface interactions at the atomic scale. Prior studies by Takahashi et al. \cite{takahashi2010molecular} and Saiz et al. \cite{saiz2014atomistic} modeled \ce{EMI-BF4} nanodroplets impacting a tungsten surface and crystalline silicon using non-reactive Lennard‑Jones (LJ) potentials to model every molecular interaction in the Large-scale Atomic/Molecular Massively Parallel Simulator (LAMMPS) \cite{thompson2022lammps}. These computational models limited the chemical fidelity by excluding bond cleavage, bond formation, and any electronic processes at the metal surface. Saiz indicates that sputtering begins around $12.7$ \unit{eV} per molecule and continues through a collision-cascade phase followed by thermal evaporation. The sputtering yield increases with impact kinetic energy and, at fixed velocity, decreases with droplet diameter due to backscattering. Saiz establishes energy to sputter yield scaling and ejection pathways for large polyatomic projectiles. In Cidoncha et al., Bendimerad built on this work by employing non-reactive simulations of a single neutral \ce{EMI-BF4} pair impacting a \ce{Au} slab with a hybrid potential: LJ terms for the projectile IL molecule and an embedded atom method (EAM) potential for \ce{Au}-\ce{Au} interactions \cite{cidoncha2022modeling}. Bendimerad mapped reflection and ionic bond dissociation thresholds across $10$ to $90$ \unit{\eV} and multiple projectile orientations without explicit modeling of chemical recombination or charge transfer. Recently, Bendimerad et al. extended this work by introducing reactive MD using the reaxFF force field \cite{van2001reaxff} to incorporate covalent bond breakage and formation \cite{bendimerad2022molecular} for \ce{EMI-BF4} projectile impacts to \ce{Au} surfaces between $10$ to $1000$ \unit{\eV}. Bendimerad reports pseudomass spectra and fragmentation channels across impact energies and molecular orientations. Bendimerad notes that neutral \ce{EMI-BF4} covalent fragmentation begins at $10$ \unit{eV}, producing fragments such as \ce{BF2}, \ce{BF3}, \ce{F}, and larger IL species. These classical MD approaches clarify kinematics and covalent fragmentation pathways, but do not include explicit electronic degrees of freedom for the incident projectile at the metal surface. As a result, these modeling efforts cannot predict fragment charge state, energy transfer mechanisms, and the recombination of radical species into neutral molecules such as \ce{HF} and \ce{H2} \cite{shaik2024characterization} that are central to SSE.

The present work overcomes these limitations by applying mixed quantum/classical (QM/MM) ab initio molecular dynamics (MD) to simulate the charge evolution of neutral \ce{EMI-BF4} impacts on gold extractor surfaces across the $10$ to $100$ \unit{\eV} kinetic energy window.

\section{Methodology}

\subsection{Computational Framework}

\begin{figure}[htb!]
    \centering
    \includegraphics[width=0.99\linewidth]{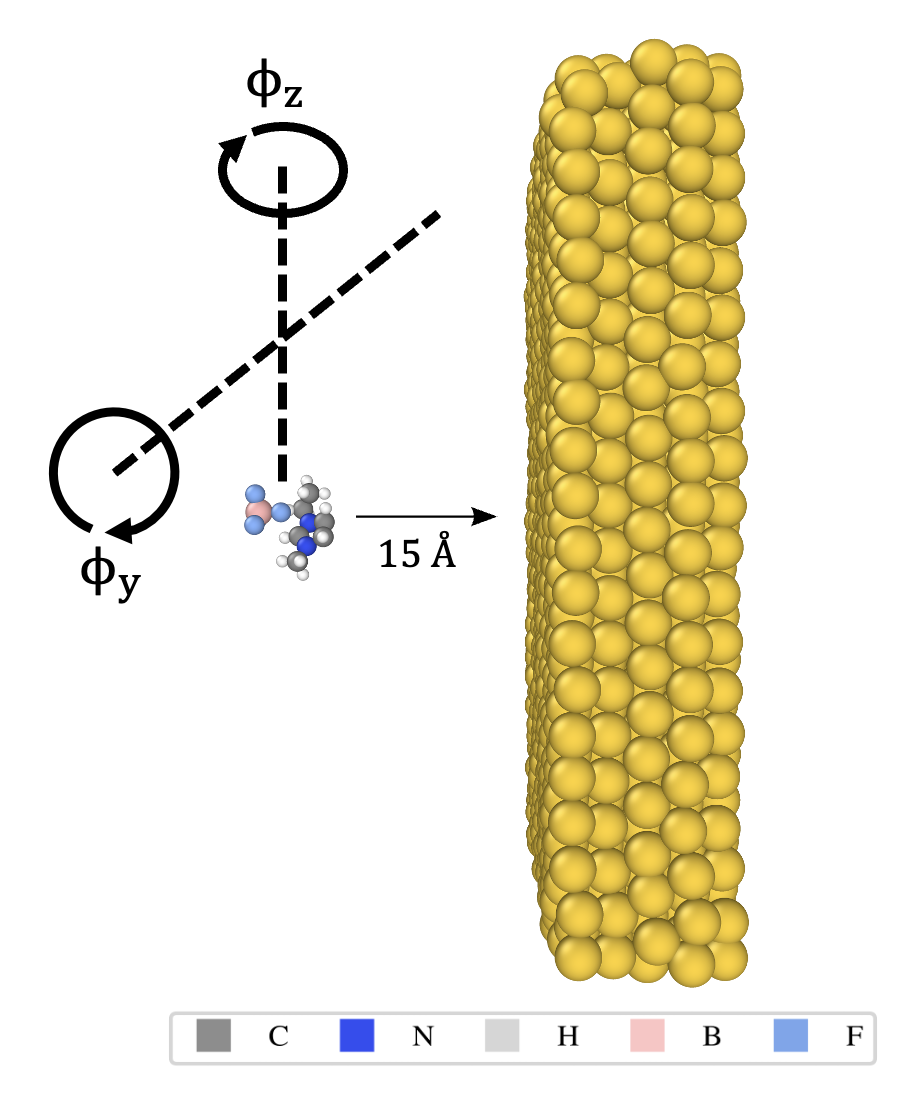}
    \caption{Simulation domain for extractor surface collisions from incident \ce{EMI-BF4} neutrals.}
    \label{fig:computational_setup}
\end{figure}

\begin{figure*}[htb!]
    \centering
    \includegraphics[width=0.9\linewidth]{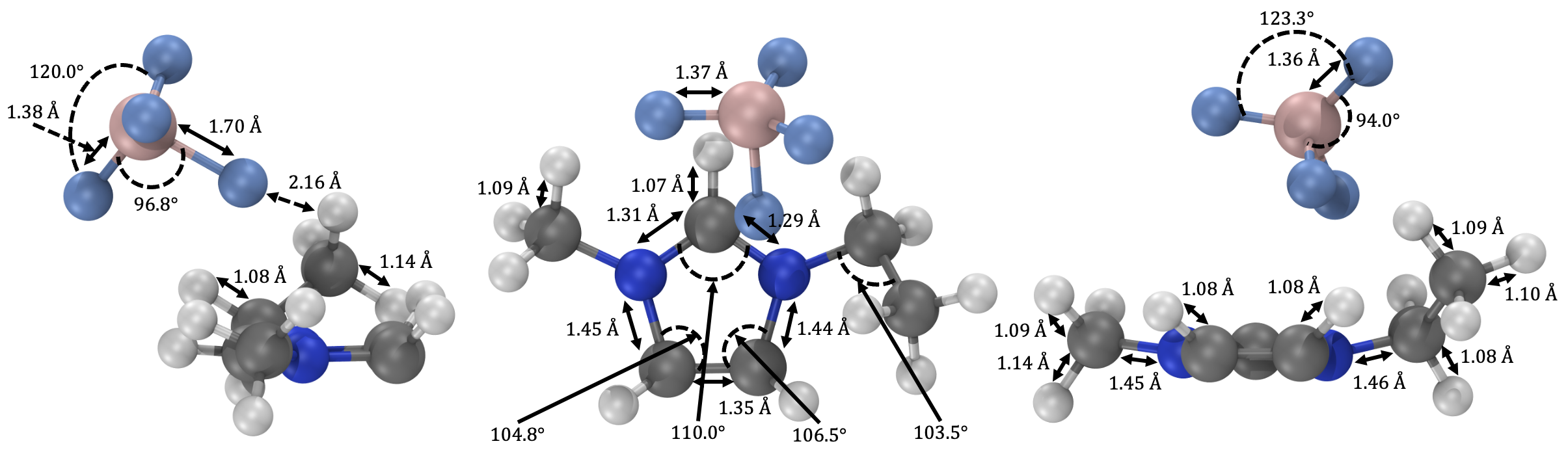}
    \caption{Molecular geometry of the DFT optimized incident \ce{EMI-BF4} projectile from the right, top, and front views.}
    \label{fig:geometry_optimized}
\end{figure*}

\begin{figure}[htb!]
    \centering
    \includegraphics[width=0.99\linewidth]{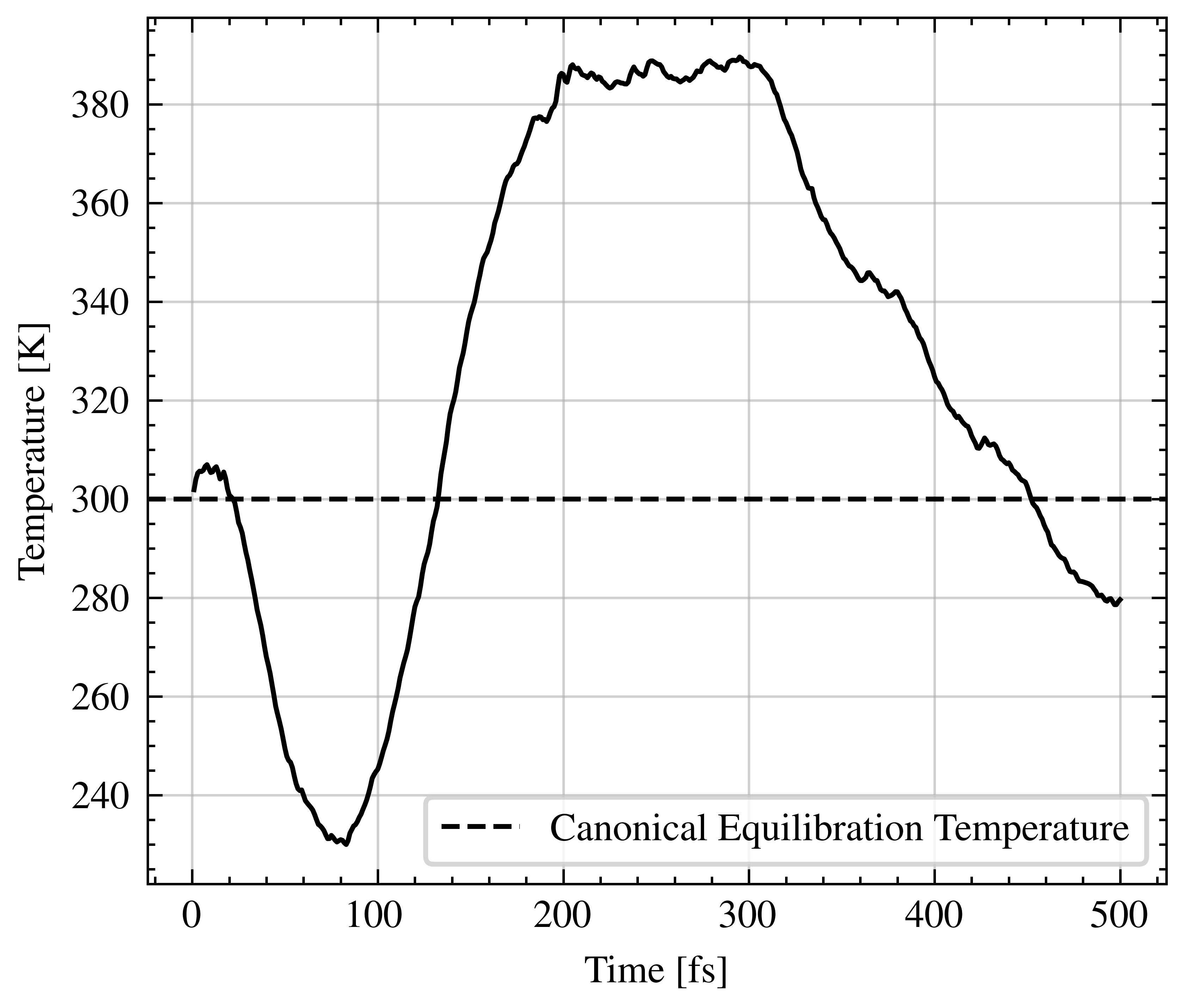}
    \caption{Equilibration temperature for the incident \ce{EMI-BF4} neutral during the canonical ensemble.}
    \label{fig:NVT}
\end{figure}

The charge‐resolved collision dynamics of \ce{EMI-BF4} with a gold extractor surface were investigated with CP2K's mixed quantum, classical (QM/MM) module \cite{kuhne2020cp2k}. CP2K’s Quickstep solver and Gaussian and augmented plane wave (GAPW) formulation is leveraged to propagate MD trajectories. CP2K's GAPW module combines localized Gaussian orbitals with uniform plane‑wave grids to achieve all‑electron accuracy for first‑row elements at moderate computational cost. Figure~\ref{fig:computational_setup} depicts the initial configuration of the simulation cell bounded by $75 \times 75 \times 75$ {\AA}$^3$ vacuum to suppress image interactions in the microcanonical ensemble. The neutral species \ce{EMI-BF4} is geometry optimized as shown in Figure~\ref{fig:geometry_optimized}, quickly temperature equilibrated to $300$ \unit{\K} over $500$ \unit{\fs} as depicted in Figure~\ref{fig:NVT} to thermalize the system while retaining the optimized geometry, and is placed $15$ {\AA} orthogonally away from the center of a frozen \ce{Au} slab measuring $16 \times 16 \times 4$ {\AA}$^3$, totaling $2000$ atoms. To benchmark and validate the optimized \ce{EMI-BF4} structure used as the incident neutral, key intramolecular metrics are compared against a single-crystal X-ray structure of \ce{EMI-BF4} collected at $173$ \unit{\K} \cite{choudhury2005situ} as described in Table~\ref{tab:geometry}. Density functional theory (DFT) optimizations predict three of the four \ce{B-F} bonds within $0.02$ {\AA} of experimental crystallography measurements, whereas one of the \ce{B-F} bonds is elongated by $0.304$ {\AA}. The single elongated \ce{B-F} likely stems from directional \ce{C-H}$\cdots$\ce{F} bonding from the \ce{EMI+} cation, which polarizes the \ce{BF4-} anion toward one \ce{F} atom and lengthens its attached \ce{B-F} bond. DFT optimizations calculate the six \ce{F-B-F} angles to span a broader range in the optimized pair, $94.0$ to $123.3$$^{\circ}$, than the experimental measurements of $108.7$ to $111.1$$^{\circ}$. The imidazolium ring metrics are well reproduced in DFT and are consistent with environmental effects. The isolated contact ion pair optimized in vacuum can relax toward an anisotropic arrangement that strengthens cation to anion contact whereas the solid-state packing of the crystal constrains \ce{BF4-} closer to tetrahedral geometry. The direct comparison of DFT geometry optimization metrics to experimental crystallography validates the fragmentation pathways upon surface impact.

\setcounter{table}{0}
\begin{table}[ht]
\caption{\label{tab:geometry}
Comparison of DFT optimized \ce{EMI-BF4} geometry metrics with experimental crystallography measurements of \ce{EMI-BF4} at $173$ \unit{\K} \cite{choudhury2005situ}.}
\begin{ruledtabular}
\begingroup\renewcommand{\arraystretch}{1.15}
\newcommand{\grouprule}{\noalign{\vskip 2pt}\colrule\noalign{\vskip 2pt}}
\newcommand{\subtitlefirst}[1]{%
  \multicolumn{3}{c}{\textbf{#1}}\\%
  \grouprule
}
\newcommand{\subtitleboth}[1]{%
  \grouprule
  \multicolumn{3}{c}{\textbf{#1}}\\%
  \grouprule
}
\begin{tabular}{lcc}
\textbf{Measurement} & \textbf{DFT} & \textbf{Experimental Crystallography} \\
\colrule
\subtitlefirst{Anion Bond Lengths (\AA)}
\ce{B_1-F_1} & 1.364 & 1.376 \\
\ce{B_1-F_2} & 1.365 & 1.386 \\
\ce{B_1-F_3} & 1.378 & 1.391 \\
\ce{B_1-F_4} & 1.703 & 1.399 \\
\subtitleboth{External Anion Angles ($^\circ$)}
\ce{F_1-B_1-F_2} & 94.0  & 108.7 \\
\ce{F_1-B_1-F_3} & 96.0  & 108.8 \\
\ce{F_1-B_1-F_4} & 96.8  & 109.0 \\
\ce{F_2-B_1-F_3} & 113.9 & 109.5 \\
\ce{F_2-B_1-F_4} & 120.0 & 109.7 \\
\ce{F_3-B_1-F_4} & 123.3 & 111.1 \\
\subtitleboth{Cation Bond Lengths (\AA)}
\ce{N_1-C_2} & 1.292 & 1.330 \\
\ce{C_2-N_3} & 1.305 & 1.335 \\
\ce{N_3-C_4} & 1.346 & 1.361 \\
\ce{C_4-C_5} & 1.436 & 1.384 \\
\ce{C_5-N_1} & 1.452 & 1.390 \\
\subtitleboth{Internal Cation Angles ($^\circ$)}
\ce{C_5-N_1-C_2} & 104.8 & 106.3 \\
\ce{N_1-C_2-N_3} & 106.5 & 107.8 \\
\ce{C_2-N_3-C_4} & 109.2 & 107.8 \\
\ce{N_3-C_4-C_5} & 109.3 & 108.9 \\
\ce{C_4-C_5-N_1} & 110.0 & 109.3 \\
\end{tabular}
\endgroup
\end{ruledtabular}
\begin{flushleft}
\end{flushleft}
\end{table}

MD trajectories are propagated in the microcanonical ensemble to a maximum of $2$ \unit{\ps} using an adaptive timestep integrator. Integration begins with the timestep $\Delta t = 1$ \unit{\fs} and is recursively halved whenever any atomic displacement exceeds $0.1$ {\AA} \cite{marx2009ab}. Electrostatics are evaluated with the Martyna-Tuckerman Poisson solver \cite{martyna1999reciprocal} under fully nonperiodic boundary conditions, eliminating artifacts from spurious periodic images.

In each trajectory, the incident neutral projectile is assigned a translational velocity in the positive $\hat{x}$ direction that reproduces a prescribed center-of-mass kinetic energy, $KE_{i}$. For a rigid body of molar mass $M$, the classical relation

\begin{equation}
    \begin{aligned}
        v_{CoM} = \sqrt{\frac{2KE_{i}}{m}} \text{ with } m = \frac{M}{N_A}
    \end{aligned}
    \label{equation:com}
\end{equation}

\noindent converts $KE_{i}$ in \unit{\joule} to the laboratory-frame speed $v_{CoM}$ in ${m}/{s}$ where $N_A$ is the Avogadro constant. The velocity is subsequently expressed in \si{\bohr}$/$\si{\autime}. Only the component orthogonal to the surface normal, $\hat{x}$, is augmented, thereby isolating impact energy as the sole control parameter.

Atomic coordinates extracted from the equilibrated NVT snapshot are rotated so that the molecular dipole aligns with a target incidence vector $V_{i}$. This rotation employs Rodrigues’ formula, which converts an axis-angle pair into the rotation matrix

\begin{equation}
    \begin{aligned}
        R = I + \sin{\theta}K + \left(1-\cos{\theta}\right)K^2 
    \end{aligned}
    \label{equation:com}
\end{equation}

\begin{equation}
    \begin{aligned}
        K = \begin{bmatrix}
        0 & -k_z & k_y \\
        k_z & 0 & -k_x \\
        -k_y & k_x & 0
        \end{bmatrix}
    \end{aligned}
    \label{equation:com}
\end{equation}

\noindent where $k_d = {u}/{\| u \|}$  is the unit vector along $u = V_{0} \times V_{i}$ and $\theta = \arccos{{V_{0} \cdot V_{i}}/{\|V_{0}\| \|V_{i}\|}}$.

\begin{figure}[htb!]
    \centering
    \includegraphics[width=0.99\linewidth]{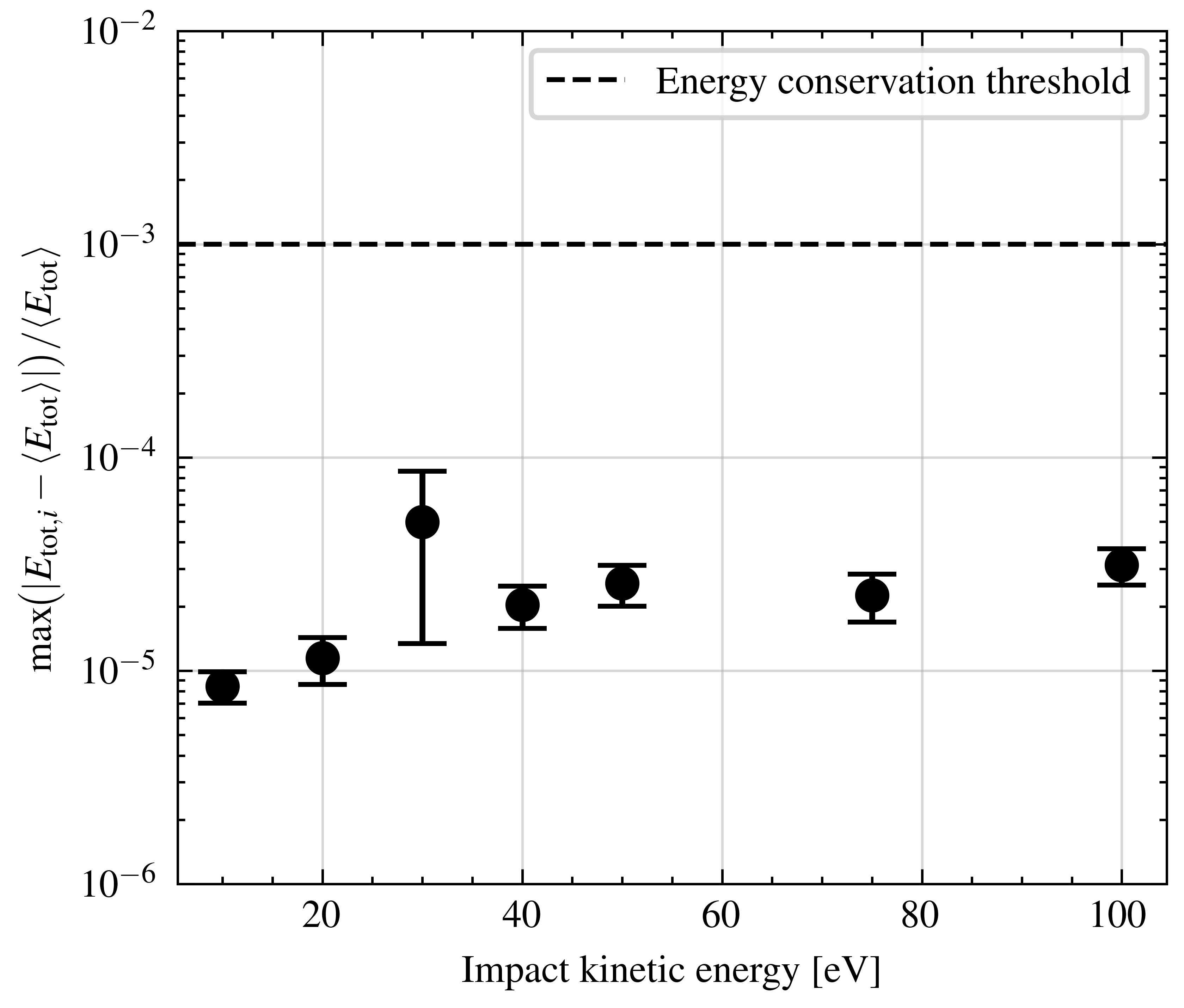}
    \caption{Maximum relative deviation of the total energy in microcanonical trajectories across impact kinetic energies.}
    \label{fig:energy_conservation}
\end{figure}

To verify numerical stability and rule out spurious heating in the microcanonical trajectories, the total QM/MM energy, nuclear kinetic, Kohn-Sham electronic, and QM-MM interactions terms, are monitored to ensure conservation. For each trajectory, the worst‑case relative energy deviation is computed using Equation~\ref{eq:validation}

\begin{equation}
    \begin{aligned}
        \delta E_{max} = \frac{\mathrm{max}_i\left(E_{tot,i} - \left<E_{tot}\right>\right)}{\left<E_{tot}\right>}
    \end{aligned}
    \label{eq:validation}
\end{equation}

\noindent where $E_{tot,i}$ is the instantaneous total energy at MD step $i$ and $\left<E_{tot}\right>$ is the time-average over the microcanonical trajectory. A $\delta E_{max}$ threshold of $10^-3$ is selected as the acceptance criterion for energy conservation, meaning that values below this boundary indicate that integration error and electronic self-consistency residuals are negligible as compared with energy exchanges on the scale of \unit{\eV}. Figure~\ref{fig:energy_conservation} summarizes $\delta E_{max}$ as a function of impact kinetic energy. Across the $10$ to $100$ \unit{\eV} window, all trajectories satisfy the conservation criterion by more than an order of magnitude. The error bars capture the variability across repeated trajectories with different orientations at a given impact kinetic energy.

\subsection{Parameter Space}

Energy-resolved tandem mass spectrometry shows that covalent fragmentation of \ce{EMI-BF4} monomers begins at less than $5$ \unit{\eV} and reaches full bond cleavage by $35$ to $45$ \unit{\eV} for gas-phase ions colliding with inert \ce{N2} \cite{bell2024experimental}. The parameter space of this study targets the collision energy window where IL projectiles fragment on impact but do not yet produce significant sputtering or SEE. Accordingly, prior nonreactive and reactive MD campaigns have confined the simulation domain to a maximum of $100$ \unit{\eV} impacts \cite{bendimerad2022molecular, shaik2024characterization}. Guided by these experimental findings, we restrict this study’s MD campaign to neutral \ce{EMI-BF4} impacts with \ce{Au}. The parameter matrix of impact kinetic energies and orientation vectors used to sample the fragmentation kinetics appears in Table~\ref{tab:parameter_space}.

\begin{table}[ht]
\caption{\label{tab:parameter_space}Ranges of impact kinetic energy and Euler angles ($\phi_y$, $\phi_z$) used to initialize each microcanonical trajectory.}
\begin{ruledtabular}
\begin{tabular}{lccccc}
Parameter &
Min. Value &
Max. Value &
N &
Units\\
\hline \\ [-1.75ex]
Impact Kinetic Energy & 10 & 100 & 7 & \unit{\eV} \\
$\phi_y$ & -90 & 180 & 4 & degrees \\
$\phi_z$ & -90 & 90 & 2 & degrees \\
Total &  &  & 42 &  \\
\end{tabular}
\end{ruledtabular}
\begin{flushleft}
\end{flushleft}
\end{table}

\subsection{Definition of Quantum Region}

The quantum (QM) partition of the simulation cell is defined as the complete $24$-atom \ce{EMI-BF4} projectile and every subsequent fragmented species during the collision trajectory. By keeping all reactive bonds inside the QM region, the simulation avoids artificially linked atoms and suppresses charge leakage across the QM/MM boundary. The QM region is housed in a $25 \times 25 \times 25$ {\AA}$^3$ GAPW box that is recentered on the instantaneous center of mass of the largest molecule after every MD step. This ensures a minimum of $5$ {\AA} buffer between any QM atom and the grid edge, thereby eliminating aliasing artifacts. Electronic interactions in the QM region are described with the GAPW implementation in CP2K \cite{vandevondele2005quickstep}. Each \ce{C}, \ce{N}, \ce{H}, \ce{B}, and \ce{F} atom carries the aug-cc-pVTZ basis set. Its uncontracted Gaussian primitives are described in Equation~\ref{equation:gaussian_primitives}

\begin{equation}
    \begin{aligned}
       \chi_{\mu}\left(r_e\right) = \sum_{k_p} d_{\mu k_p} \left(\frac{\alpha_{k_p}}{\pi}\right)^{\frac{3}{4}} \exp(-\alpha_{k_p} \left|r_e-R_{\mu}\right|^2)
    \end{aligned}
    \label{equation:gaussian_primitives}
\end{equation}

\noindent where $\chi_{\mu}\left(r_e\right)$ is the contracted Gaussian-type basis functions associated with atomic center $\mu$ and evaluated at the electron position $r_e$, the sum is evaluated over $k_p$ for all primitive Gaussians that form the contraction, $d_{\mu k_p}$ represents the contraction coefficients, $\alpha_{k_p}$ are the primitive exponents, and $R_{\mu}$ represents the position vector of the nucleus on which basis function $\mu$ is centered. The aug-cc-pVTZ basis set supply both diffuse and polarization functions that are essential for capturing long-range charge transfer in ionic-liquid fragments \cite{hunt2018quantum}. Exchange-correlation energies are evaluated with the $\omega$B97x-V range-separated hybrid functional \cite{mardirossian2014omegab97x} with energy decomposition as given in Equation~\ref{equation:energy_decomposition}.

\begin{equation}
    \begin{aligned}
       E_{XC} = aE_X^{HF,LR}\left(\omega\right) + \left(1-a\right)E_X^{GGA,LR}\left(\omega\right)\\ + E_X^{GGA,SR}\left(\omega\right) + E_C^{GGA} + E_C^{NL}
    \end{aligned}
    \label{equation:energy_decomposition}
\end{equation}

\noindent where $E_X^{HF,LR}\left(\omega\right)$ is the long-range Hartree-Fock exchange evaluated with a range-separated Coulomb kernel set by $\omega$, $E_X^{GGA,LR}\left(\omega\right)$ and $E_X^{GGA,SR}\left(\omega\right)$ are the long and short range components of the semilocal generalized gradient approximation (GGA) exchange, $E_C^{GGA}$ is the semilocal GGA correlation, $E_C^{NL}$ is the nonlocal correlation term such that the  exact‑exchange mixing coefficient $a = 0.167$ is supplemented by the rVV10 non-local dispersion kernel to recover London forces with plane-wave efficiency \cite{vydrov2010nonlocal}, and $\omega$ is the range‑separation parameter controlling the long and short range partitions . Exact exchange integrals are accelerated by the Auxiliary Density Matrix Method (ADMM-Q), which reduces hybrid overhead by projecting onto a compact auxiliary basis set while retaining \si{\milli\hartree} accuracy \cite{guidon2010auxiliary}.

\subsection{Definition of Classical Region}

\begin{figure}[htb!]
    \centering
    \includegraphics[width=0.99\linewidth]{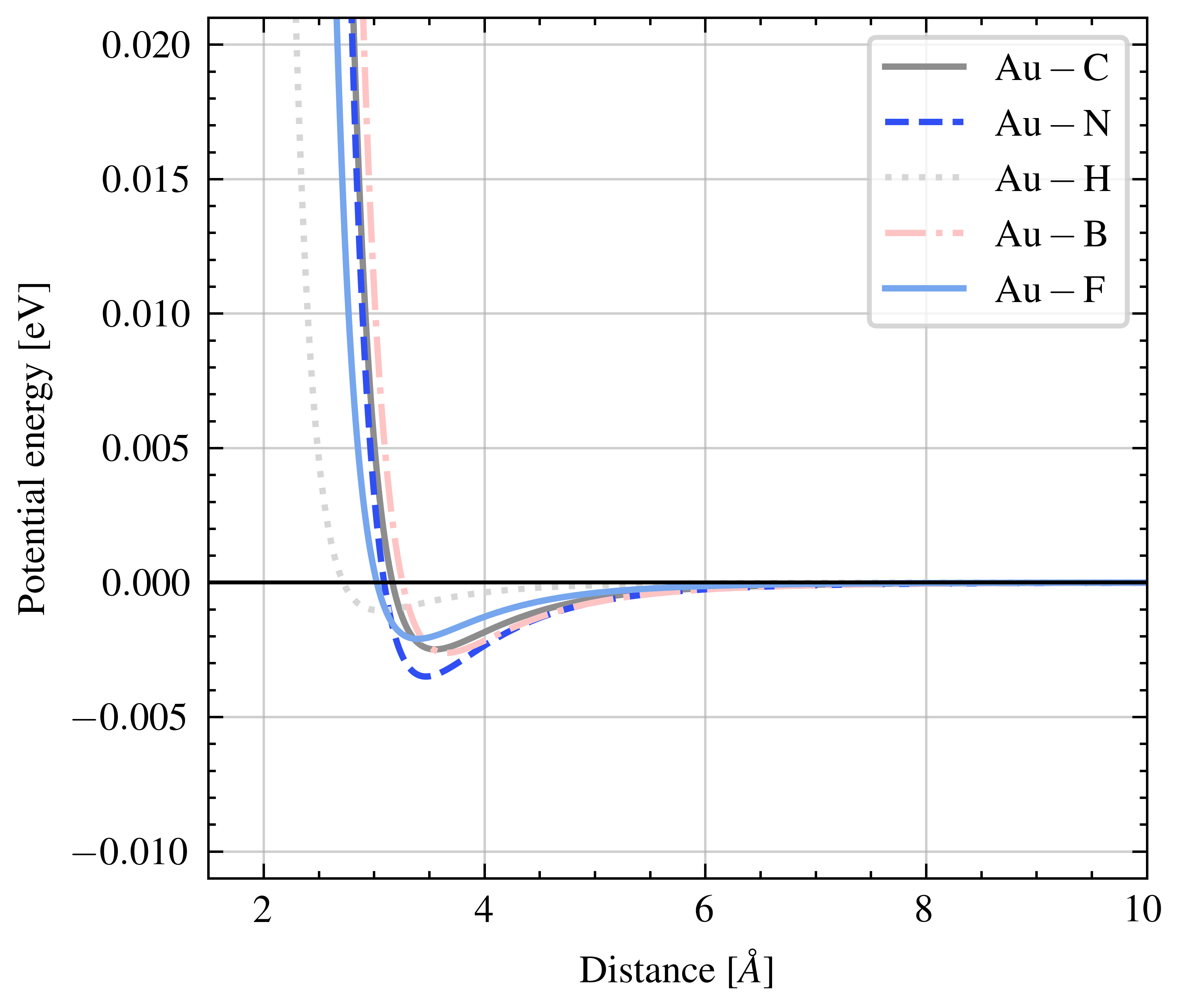}
    \caption{Lennard-Jones pair potentials for short-range interactions between \ce{Au} and QM atoms.}
    \label{fig:lj_params}
\end{figure}

The classical (MM) partition represents a rigid \ce{Au} extractor grid together with the $75 \times 75 \times 75$ {\AA}$^3$ electrostatic vacuum beyond the QM box. \ce{Au}-\ce{Au} cohesion is described by the Foiles-Baskes-Daw EAM potential \cite{foiles1986embedded}, the parameterization of which reproduces the lattice constant, cohesive energy, and elastic modulus of \ce{Au} within experimental uncertainty. Because the collision energies examined here, $10$ to $100$ \unit{\eV}, are well below the sputtering threshold, the slab is held rigid to transit momentum accurately while preventing spurious hydrogen embedding. Pauli repulsion and dispersion between the QM fragments and surface atoms are captured through LJ cross terms as defined by Equation~\ref{eq:lj_equation}

\begin{equation}
    \begin{aligned}
       U_{ij} = 4\epsilon_{ij} \left[\left(\frac{\sigma_{ij}}{r_{LJ}}\right)^{12} + \left(\frac{\sigma_{ij}}{r_{LJ}}\right)^{6}\right]
    \end{aligned}
    \label{eq:lj_equation}
\end{equation}

\noindent which defines the $12$-$6$ LJ potential applied to each \ce{Au}-$j$ pair, where $r_{LJ}$ is the interatomic separation. Arithmetic mixing of Lennard-Jones parameters following the Lorentz-Berthelot rules \cite{lorentz1881ueber, berthelot1898melange} is leveraged to calculate the well depths, $\epsilon$, and cutoff parameters, $\sigma$ for each \ce{Au}-$j$ pair using elementwise site values for \ce{N-N}, \ce{C-C}, \ce{H-H}, \ce{B-B}, and \ce{F-F} \cite{de2002computational} as well as \ce{Au-Au} \cite{huber2012thermal}, as described in Equations~\ref{equation:sigma_mixing_rules} and \ref{equation:epsilon_mixing_rules}.

\begin{equation}
    \begin{aligned}
    \sigma_{ij}=\frac{\sigma_i+\sigma_j}{2}
    \end{aligned}
    \label{equation:sigma_mixing_rules}
\end{equation}

\begin{equation}
    \begin{aligned}
    \varepsilon_{ij}=\sqrt{\varepsilon_i\varepsilon_j}
    \end{aligned}
    \label{equation:epsilon_mixing_rules}
\end{equation}

The well depths and cutoff parameters were calibrated to reproduce adhesion energies of \ce{EMI-BF4} on \ce{Au} and are depicted in Figure~\ref{fig:lj_params} and summarized in Table~\ref{tab:lj_params}.  

\begin{table}[ht]
\caption{\label{tab:lj_params} Lennard-Jones well depths $\epsilon_{ij}$ and cutoff parameters $\sigma_{ij}$ for non-bonded interactions between surface, \ce{Au}, atoms and the constituent atoms of the \ce{EMI-BF4} propellant.}
\begin{ruledtabular}
\begin{tabular}{lcc}
Pair &
$\epsilon_{ij}$ (\unit{\eV}) &
$\sigma_{ij}$ ({\AA}) \\
\hline \\ [-1.75ex]
\ce{Au}-\ce{C} & 0.00249 & 3.167 \\
\ce{Au}-\ce{N} & 0.00350 & 3.092 \\
\ce{Au}-\ce{H} & 0.00104 & 2.722 \\
\ce{Au}-\ce{B} & 0.00262 & 3.258 \\
\ce{Au}-\ce{F} & 0.00210 & 3.0261 \\
\end{tabular}
\end{ruledtabular}
\begin{flushleft}
\end{flushleft}
\end{table}

\subsection{Fragmentation Criterion} \label{sec:frag_criterion}

Bond cleavage and formation is detected by tracking whether all atoms of the neutral ion cluster projectile remain connected throughout the microcanonical trajectory. At the $k^{th}$ microcanonical frame, the instantaneous Cartesian position of atom $i$ such that $1 \le i \le N$ is defined as $r_i^{\left(k\right)}$. Pairwise interatomic separations are computed using Equation~\ref{distance_matrix}.


\begin{equation}
    d_{ij}^{(k)}=\bigl\lVert\mathbf r_i^{(k)}-\mathbf r_j^{(k)}\bigr\rVert ,
    \qquad 1\le i<j\le N
    \label{distance_matrix}
\end{equation}

These distances are converted into a bond connectivity matrix defined by Equation~\ref{bond_matrix}

\begin{equation}
    A_{ij}^{(k)}=\begin{cases}
                    1,& d_{ij}^{(k)}\le d_b,\\
                    0,& \text{otherwise}
                \end{cases}
    \label{bond_matrix}
\end{equation}

\noindent where a bond length threshold of $d_b = 5$ {\AA} is selected. For each frame $k$, the connectivity matrix $A^{(k)}$ defines the undirected graph $G^{\left(k\right)} = \left(V, A^{\left(k\right)}\right)$ with a vertex set $V = \{1, ... , N\}$. Atoms are assigned to the same fragment when they are connected through at least one continuous path of pairwise connections. The number of disconnected groups in $G^{\left(k\right)}$, denoted $C^{\left(k\right)}$, is interpreted as the fragment count at frame $k$. The microcanonical trajectory is treated as intact when $C^{\left(k\right)} = 1$. The fragmentation index $k_{\mathrm{frag}}$ is described in Equation~\ref{eq:frag_index}


\begin{equation}
    k_{\mathrm{frag}} = \min\left(k>k_0 : C^{(k)}>1\right)
    \label{eq:frag_index}
\end{equation}

\noindent which marks the first post-equilibration frame at which the neutral species has irreversibly split into two or more fragments. 

\subsection{Charge Analysis Protocol}

Time-resolved atomic charges are extracted to quantify electron exchange between the projectile and its constituent atoms during impact. All charge analyses were performed with Multiwfn \cite{lu2012multiwfn}. Multiwfn is a wavefunction analyzer that supplies a variety of population analysis schemes. This study leverages the restrained electrostatic potential (RESP) model because it reproduces intermolecular electrostatics while damping excessive charge transfer on buried atoms \cite{bayly1993well}. For each timestep, the partial charges $\left\{q_{A}\right\}$ are obtained by minimizing the RESP objective function described in Equation~\ref{eq:resp}

\begin{equation}
    \chi^2 = \sum_{i} w_i \left[V_i^{QM} - \sum_{A} \frac{q_{A}}{r_{A_i}}\right]^2 + \sum_{A} \lambda_A \left(q_A - q_A^0\right)^2
    \label{eq:resp}
\end{equation}

\noindent where $V_i^{QM}$ is the ab initio electrostatic potential at grid point $i$, $w_i$ is the uniform weight $\left(0.001 \text{ a.u.}\right)$, $r_{A_i}$ is the distance from atom $A$ to point $i$, and the second term applies a hyperbolic restraint $\lambda_A = 5 \times 10^{-4}$ a.u. that prevents overpolarization. The grid spacing is fixed to $0.3$ {\AA}, and points lying within $1.4$ {\AA} of any atomic center are excluded. 

Connectivity matrices generated from Section~\ref{sec:frag_criterion} tag every atom with a fragment label $k$. The net charge of fragment $k$ at time $t$ is described by Equation~\ref{eq:net_charge}. Charges are sampled every timestep from the moment of microcanonical ensemble initiation until $2$ ps after impact. 

\begin{equation}
    Q_{k}\left(t\right) = \sum_{A \in k}q_{A}\left(t\right)
    \label{eq:net_charge}
\end{equation}

\begin{figure}[htb!]
    \centering
    \includegraphics[width=0.99\linewidth]{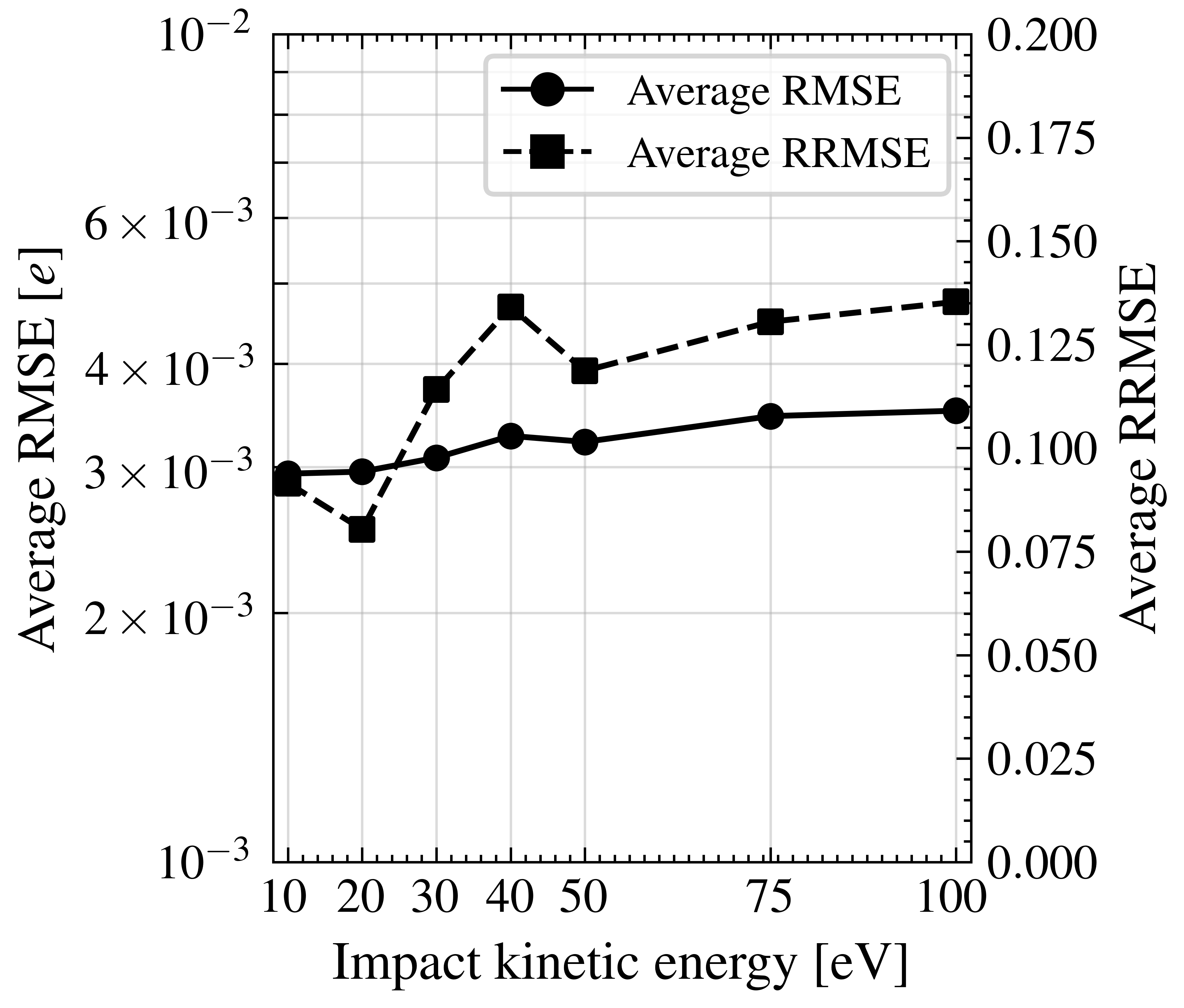}
    \caption{Root mean square error and relative root mean square error averaged across impact energies derived from the RESP fit.}
    \label{fig:charge_error}
\end{figure}

To map time-dependent RESP charges to integer charge states observed experimentally, fragments charges are binned using a small dead-band around $0$ such that fragments with $Q_k(t) \le -0.125e$ are labeled as anions, fragments with $Q_k(t) \ge +0.125e$ are labeled as cations, and fragments with $-0.125e \le Q_k(t) \le +0.125e$ are labeled as neutrals. This tolerance is intentionally larger than typical method, grid, and basis set dependent variability in partial atomic charges and avoids misclassifying weakly polarized fragments \cite{graham2023back, geiger2025secondary, hofheins2025electrospray}. To verify that the charge assignments are well-conditioned, Figure~\ref{fig:charge_error} reports the root mean square error (RMSE) and relative root mean square error (RRMSE) versus impact energy of the RESP fit. Both the absolute RMSE and RRMSE remain small and weakly dependent across $10$ to $100$ \unit{\eV}, supporting the stability of the charge analysis. 

\section{Results} \label{sec:results}

\begin{figure*}[htb!]
    \centering
    \includegraphics[width=1\linewidth]{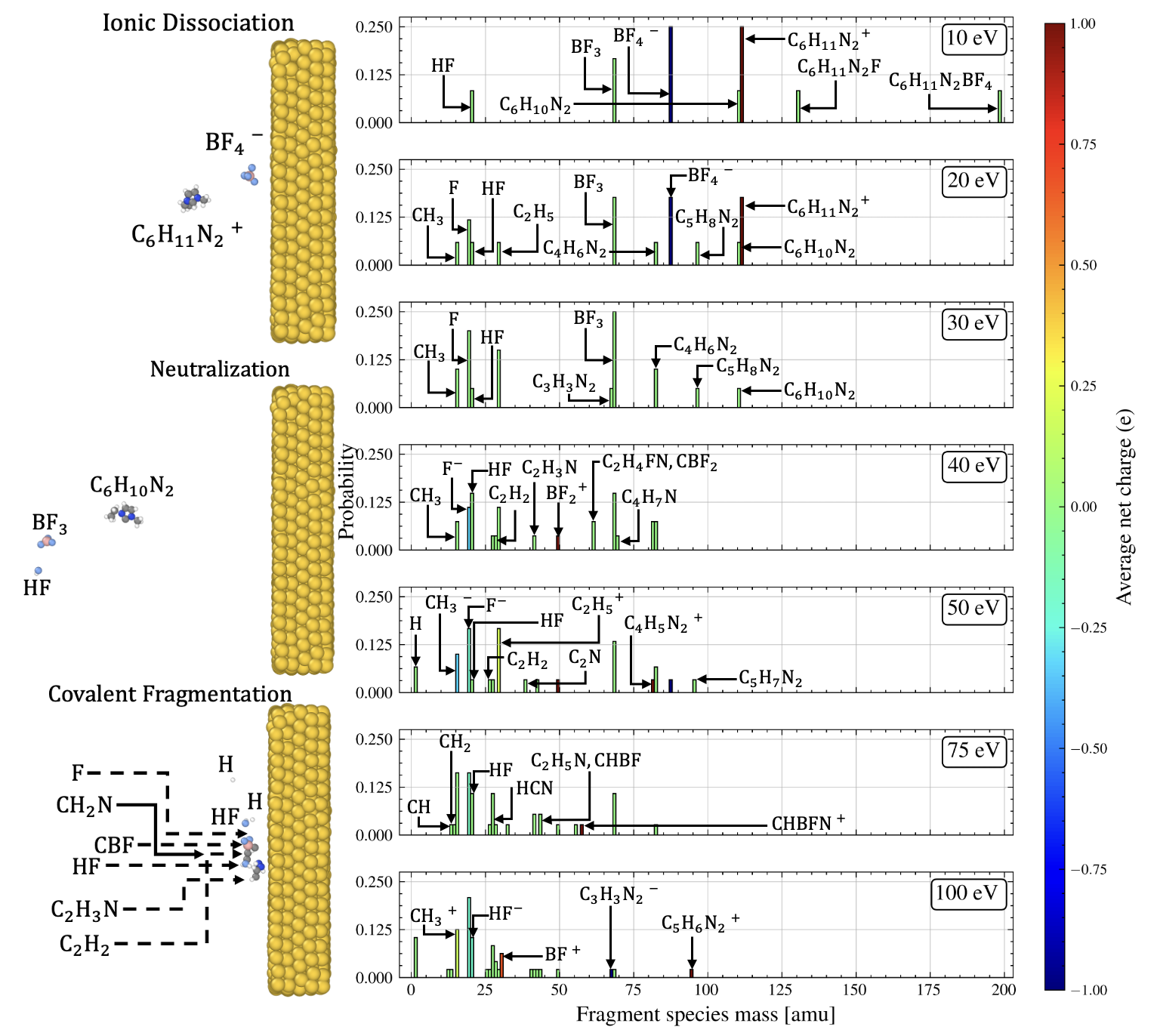}
    \caption{Probability density of fragment species mass for impact kinetic energies from $10$ to $100$ \unit{\eV}.}
    \label{fig:charge_distribution}
\end{figure*}

The surface collision ensemble reveals a clear progression of fragmentation pathways for neutral \ce{EMI-BF4} projectiles colliding with an \ce{Au} extractor surface across the $10$ to $100$ \unit{\eV} impact window. Figure~\ref{fig:charge_distribution} depicts the mass spectra colored by net fragment charge. Low energy collisions, $10$ to $20$ \unit{\eV}, are dominated by ionic dissociation of the neutral pair into the canonical IL constituents, the \ce{EMI+} cation and \ce{BF4-} anion. At intermediate energy, between $30$ to $40$ \unit{\eV}, the spectra develops strong neutralization channels, with prominent neutral products consistent with \ce{BF3} and \ce{HF} alongside a largely neutralized imidazolium backbone, \ce{C6H10N2}. Light charged fragments also appear, \ce{F-} and \ce{BF2+} within this impact energy regime for less than $25\%$ of simulations. In the high energy regime, greater than $50$ \unit{\eV}, the mass spectra distributions broaden substantially due to covalent fragmentation and become rich in light fragments with a mixture of neutral and charged products: \ce{CH3+}, \ce{CH3-}, \ce{F-}, \ce{HF-}, \ce{C2H5+}, \ce{BF+}, \ce{CHBFN+}, \ce{C3H3N2-}, \ce{C4H6N2+}, \ce{BF4-}, and \ce{C5H6N2+} as summarized in Table~\ref{tab:secondary-ion-detection}. This energy-ordered sequence aligns with facility observations that surface collisions begin to emit electrons and charged molecular secondaries once the SEE and SIE limits of $0.44$ ${eV}/{u}$ and $1$ ${eV}/{u}$ \cite{uchizono2021role}. The highest simulated energy of $100$ \unit{\eV} places our surface impact simulations in the near-onset regime for secondary production at metallic targets, implying that this limit may be lower due to the number of charged species generated below this threshold. The net charges reported in Figure~\ref{fig:charge_distribution} rests on RESP-fitted partial charges computed at every MD timestep. 

\begin{table}[ht]
\caption{\label{tab:secondary-ion-detection}
Secondary ion species detected in the microcanonical QM/MM trajectories.}
\begingroup
\renewcommand{\thefootnote}{\alph{footnote}}%
\setcounter{footnote}{0}%
\setlength{\tabcolsep}{6pt}%
\renewcommand{\arraystretch}{1.15}%
\begin{ruledtabular}
\begin{tabular}{@{}r l c@{}}
\textbf{Mass (amu)} & \textbf{Species} & \textbf{Reference Detection} \\
\colrule
15 & \ce{CH3+}, \ce{CH3-} & \citenum{gunster2008time}, \citenum{bundaleski2013ion}, \citenum{hofheins2025thinfilms} \\
19 & \ce{F-} & \citenum{van1999negative}, \citenum{gunster2008time}, \citenum{bundaleski2013ion}, \citenum{hofheins2025electrospray} \\
20 & \ce{HF-} & \citenum{van1999negative}, \citenum{gunster2008time}, \citenum{bundaleski2013ion}, \citenum{hofheins2025electrospray} \\
29 & \ce{C2H5+} & \citenum{gunster2008time}, \citenum{bundaleski2013ion}, \citenum{bell2024experimental}, \citenum{hofheins2025electrospray} \\
30 & \ce{BF+} & \citenum{gunster2008time}, \citenum{bell2024experimental} \\
49 & \ce{BF2+} & \citenum{bell2024experimental} \\
57 & \ce{CHBFN+} & \citenum{hofheins2025electrospray} \\
67 & \ce{C3H3N2-} & \citenum{bell2024experimental}, \citenum{hofheins2025electrospray}, \citenum{hofheins2025thinfilms} \\
81 & \ce{C4H5N2+} & \citenum{hofheins2025thinfilms} \\
87 & \ce{BF4-} & \citenum{van1999negative}, \citenum{bundaleski2013ion}, \citenum{bell2024experimental}, \citenum{hofheins2025thinfilms} \\
94 & \ce{C5H6N2+} & \citenum{bundaleski2013ion}, \citenum{hofheins2025thinfilms} \\
111 & \ce{C6H11N2+} & \citenum{gunster2008time}, \citenum{bundaleski2013ion}, \citenum{bell2024experimental}, \citenum{hofheins2025electrospray}, \citenum{hofheins2025thinfilms} \\
\end{tabular}
\end{ruledtabular}
\endgroup
\end{table}

\begin{figure}[htb!]
    \centering
    \includegraphics[width=0.99\linewidth]{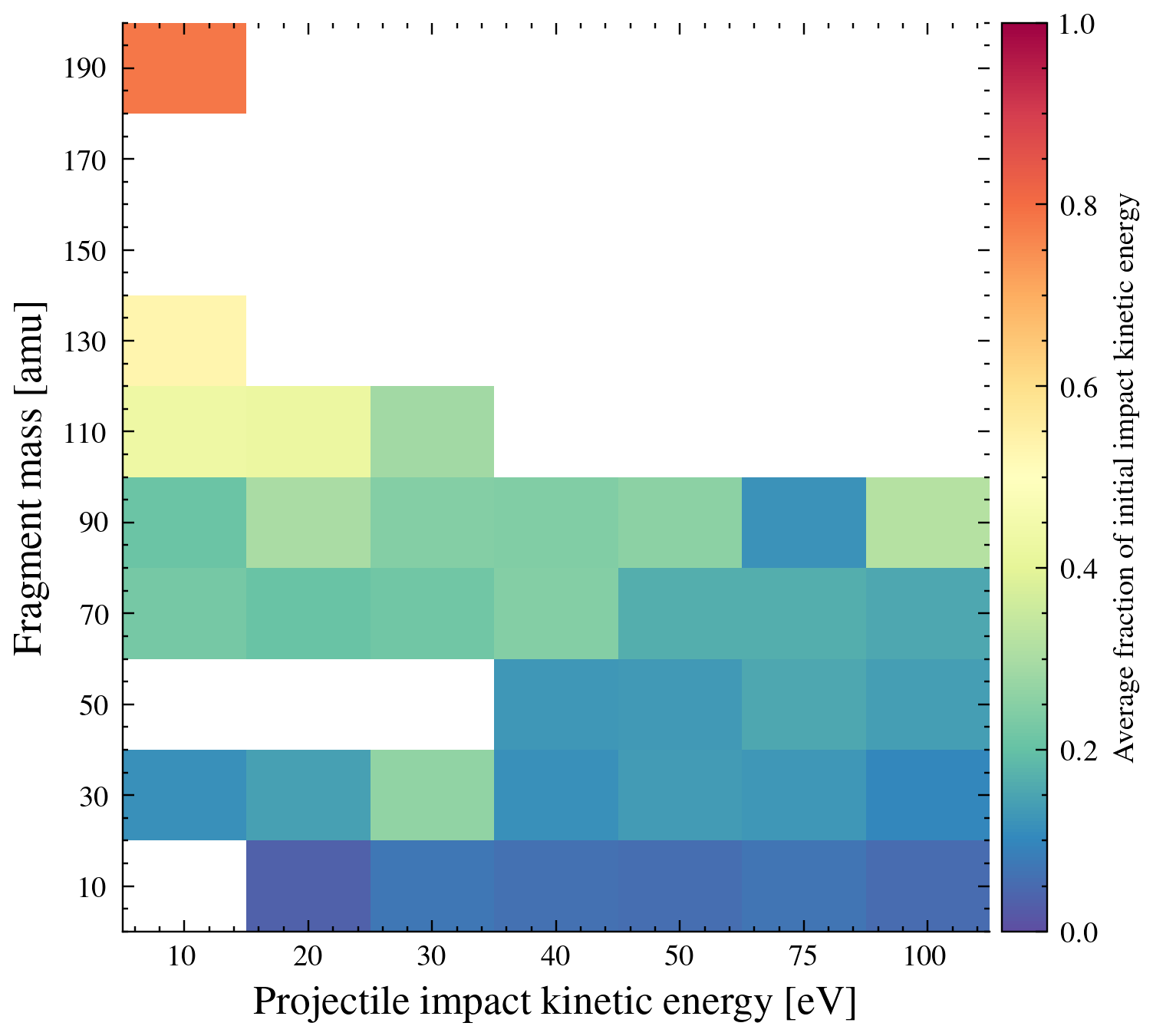}
    \caption{Mass-resolved energy partition fraction as a function of initial projectile impact energy.}
    \label{fig:impactKEDistribution}
\end{figure}

Figure~\ref{fig:impactKEDistribution} quantifies how the momentum and energy are partitioned among these products by reporting the fraction of initial impact kinetic energy that each fragment carries after collision. At $10$ to $20$ \unit{\eV}, a small number of heavy fragments, most interestingly the transient metastable species \ce{C6H11N2F} and recombined neutral \ce{EMI-BF4} pairs, carry most of the available kinetic energy. These outcomes are consistent with the ionic dissociation regime, where fragmented species dissipate less than half of the impact energy into the surface due to short-range Coulombic repulsion. Above $50$ \unit{\eV}, covalent fragmentation becomes prevalent such that the impact energy is distributed over many light fragments. By increasing impact energy, the breakup transitions from few-body, energy-concentrated outcomes to many-body, energy-dispersed outcomes. This distinction is astutely important for ground tests where heavy fragments that retain a large fraction of the incident energy are more likely to deliver momentum to nearby hardware and to seed SEE or SIE. On the other hand, although the cloud of many light fragments distributes this energy broadly in the high energy impact regime, the light neutral fragments can initiate a cascade of fragmentation with heavier species that add to mass error in TOF measurements \cite{lyne2023simple, natisin2021efficiency, de2025comparison}. Geiger et al. corroborates this finding through energy-swept \ce{Au} target studies that report robust charged-secondary production from electrospray plume impacts across $30$ to $1800$ \unit{\eV}, that bound the upper window of this study.


\begin{figure}[htb!]
    \centering
    \includegraphics[width=0.99\linewidth]{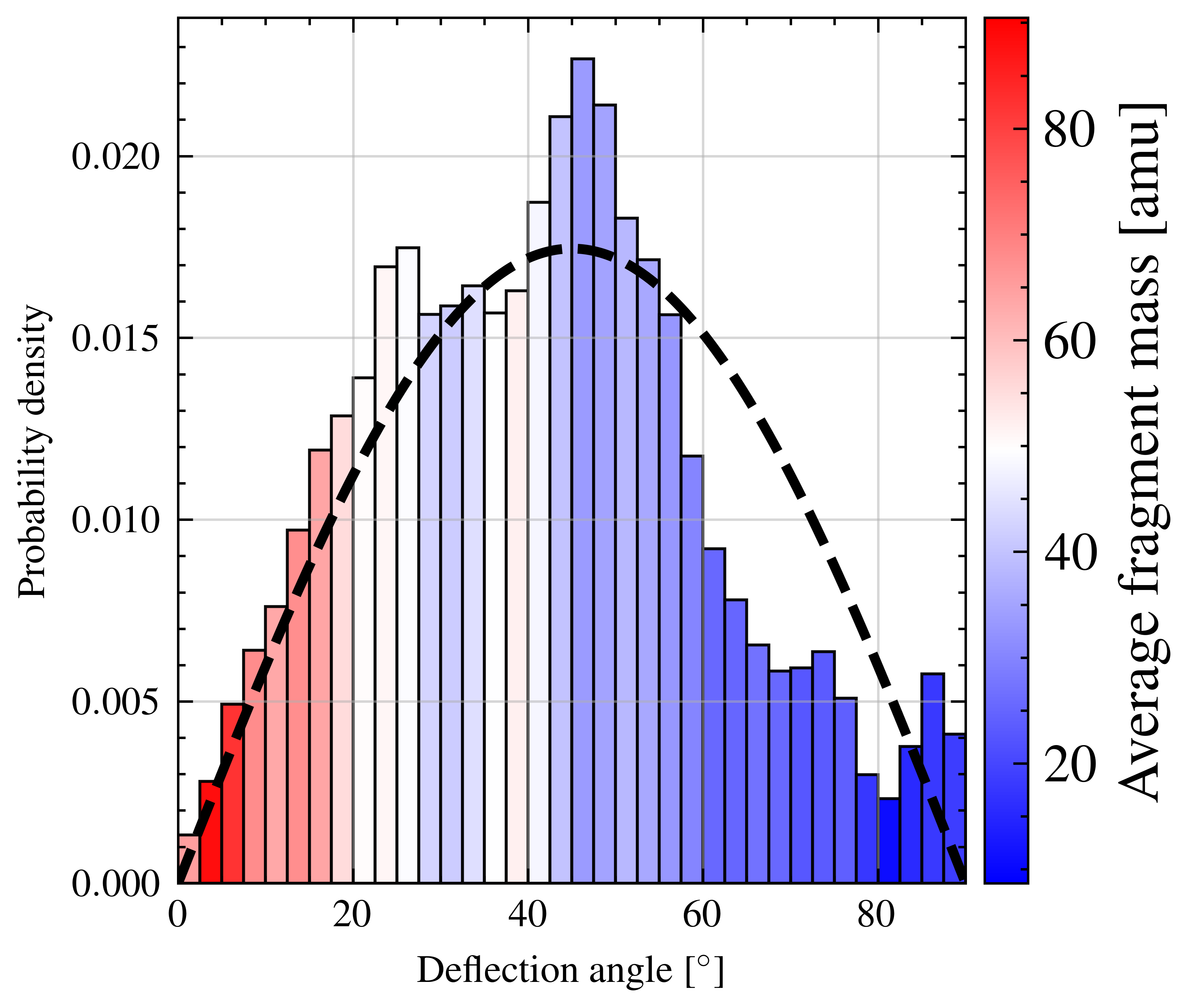}
    \caption{Fragment mass deflection angle distribution averaged across all impact energies.}
    \label{fig:deflection_angle}
\end{figure}

Figure~\ref{fig:deflection_angle} represents the deflection angle distribution of fragmented species tracked in this study. The distribution peaks at moderate angles, $40$ to $55^\circ$ and exhibits a strong mass-to-angle anti-correlation such that heavier fragments preferentially scatter closer to the forward direction, while lighter fragments populate larger deflections. The scattering behavior is approximately Lambertian and dovetails with multiscale plume studies \cite{krejci2017emission, ma2021plume, petro2022multiscale} that tie beam interception to the number and distribution of active emission sites. From this distribution, it can be noted that heavy fragments increase grid and chamber interception probability, whereas light fragment scattering increases diffuse reflection contamination that directly correlates with facility effects. 

\begin{figure}[htb!]
    \centering
    \includegraphics[width=0.99\linewidth]{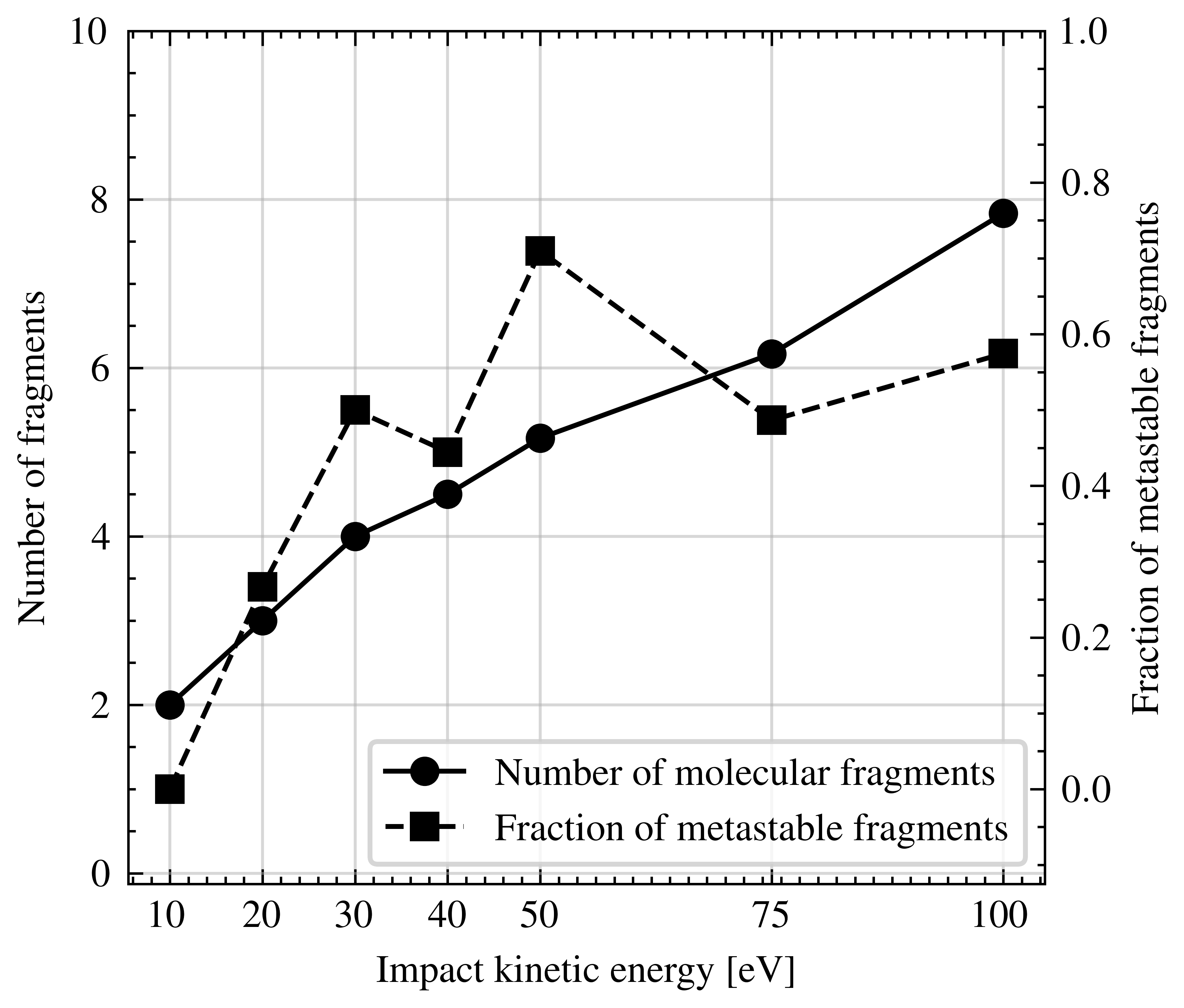}
    \caption{Transient metastable species formation as a function of impact energy.}
    \label{fig:metastable_species}
\end{figure}

The average fragment count logged during each simulation rises monotonically as impact energy increases as described in Figure~\ref{fig:metastable_species}. The fraction of transient metastable fragments, which are species that are highly reactive and appear promptly after impact, peaks near $50$ \unit{\eV} and declines at higher energies. This non-monotonicity is consistent with an intermediate energy neutralization window that readily forms volatile species such as \ce{CHBFN+} and neutrals such as \ce{HF}. The appearance of neutral byproducts at intermediate energies connects directly with in-situ RGA in electrospray facilities \cite{shaik2024characterization}, which has identified neutrals and, most importantly, \ce{HF} which is formed through: \ce{EMI-BF4} $\rightarrow$ \ce{C6H10N2} $+$ \ce{BF3} $+$\ce{HF}.

\section{Discussion}

The atomistic picture that emerges from this study is a three-stage sequence with increasing impact energy: ionic dissociation of the neutral \ce{EMI-BF4} pair in the low, $10$ to $20$ \unit{\eV}, regime, a neutralization window in the intermediate, $30$ to $40$ \unit{\eV}, regime characterized by \ce{BF4-} $\rightarrow$ \ce{BF3} $+$ \ce{HF} and a neutralized imidazolium backbone, \ce{C6H10N2}, and covalent fragmentation in the high, greater than $50$ \unit{\eV}, regime that yields many light products with mixed charge states. Placing these regimes in the context of facility measurements clarifies why charged secondaries appear for nominally sub-threshold impacts. For a neutral \ce{EMI-BF4} projectile, $100$ \unit{\eV} corresponds to approximately $0.5$ ${eV}/{u}$, falling right above the SEE onset of $0.44$ ${eV}/{u}$ and below the SIE onset of $1$ ${eV}/{u}$ \cite{uchizono2021role}. The trajectories described in this study therefore probe the near-onset regime where modest charge-secondary yields are expected and energetically plausible which is consistent with Uchizono. 

A key operational implication from this work is that neutral fragments are not benign. The intermediate energy regime where neutralization is a dominant pathway coincides with abundant neutrals that scatter broadly. Geiger et al. justifies this behavior by describing that even when the ion portion of the plume is fully suppressed by a decelerating electrode, charged secondary species are still emitted from the target under neutral bombardment \cite{geiger2025secondary}. The simulations presented in this study provide evidence that charge redistribution within the projectile prior to or during impact produces charged fragments upon collision and underscore that neutral suppression in ground tests does not eliminate SSE near the source.

The pathways presented in this study clarify diagnostic behaviors in TOF-SIMS, suppression-bias probes, and QCM measurements. TOF-SIMS diagnostics record only charged secondaries whereas the data described in Section~\ref{sec:results} predict the growth of specific charged fragments of \ce{CH3+}, \ce{CH3-}, \ce{F-}, \ce{HF-}, \ce{C2H5+}, \ce{BF+}, \ce{CHBFN+}, \ce{C3H3N2-}, \ce{C4H6N2+}, \ce{BF4-}, and \ce{C5H6N2+} with energy, while neutral channels of \ce{BF3}, \ce{HF}, and \ce{C6H10N2} remain invisible to TOF. The neutral species characterized in this study must be captured with neutral diagnostics as demonstrated by Shaik et al. using RGA \cite{shaik2024characterization}. Suppression-bias probes have been established to recover positive and negative secondary current yields that enable correction of TOF and tandem energy-analyzer measurements \cite{lyne2024quantifying}. The results presented in this study indicate that these corrections are especially important once impacts reach the neutralization and covalent fragmentation regimes where light, large angle products are abundant. QCM measurements that transition from net deposition at low impact energies to sputter dominant response at higher energies are qualitatively consistent with the energy partitioning observed in this study. At low energies, few heavy fragments retain a large share of the energy, whereas at higher energy, the same budget is distributed across many light products. 

At the plume scale, the anti-correlation between fragment mass and deflection angle recorded provides a bridge between single-impact chemistry and grid interception in arrays. Heavier fragments are biased to approximately specular reflection and, therefore, more likely to propel past extractor or accelerator hardware. Lighter fragments populate larger deflection angles, contributing to diffuse contamination that can contribute to mass error in experimental measurements \cite{lyne2023simple, natisin2021efficiency, de2025comparison}. Multiscale plume models that combine near-emitter physics with PIC transport have shown that the number and distribution of active emission sites steer beam divergence and interception \cite{hampl2022comparison, whittaker2023modeling, smith2024propagating}. By folding the energy-dependent statistics into multiscale modeling frameworks, the ability to predict facility effects at the device level is possible.

The peak of transient metastables near $50$ \unit{\eV} coincides with the onset of abundant neutral byproducts that are missed by traditional charged species diagnostics but are visible to RGA. The presence of these neutral byproducts can alter background chemistry and complicate long-duration characterization. Metastable intermediates make mass error characterization worse in single-emitter TOF measurements because prompt fragmentation and neutral formation between the source and detector reduce the charged-particle fraction sampled by TOF and retarding potential analysis (RPA). These complex surface processes bias inferred mass flow unless fragmentation models and neutral-species measurements are included. Future experimental work must develop tandem instruments that observe neutral and charged byproducts during \ce{EMI-BF4} operation and controlled plume-surface impacts.

The trajectories presented in this work employ a QM/MM partition in which the projectile and its fragments are treated quantum-mechanically while the \ce{Au} slab is classical and rigid. This model yields sufficient statistics across orientations and energies but does not allow explicit electron exchange with the metal or image-charge formation in the substrate. Therefore, fragment charges represent intra-fragment charge redistribution, and not metal-projectile electron transfer, and the conclusions made from this study are made in this context. Moreover, the $2$ ps microcanonical ensemble window captures prompt chemistry and scattering but cannot access slower surface reconstruction, diffusion, or sputtering. Future modeling efforts should include polarizable or QM layers of the metal to capture image-charge and electron-transfer pathways, probe longer windows for thermalization and desorption, and couple per-impact statistics to plume-scale models to predict array-level divergence and interception. It is recommended that experimental measurements employ suppression-bias probes during TOF and RPA characterization by applying correction procedures to de-bias spectra from backstreamed secondaries. Furthermore, it is suggested to pair TOF with tandem RGA or direct neutral diagnostics to inventory neutral byproducts that dominate the intermediate impact window and perform energy-swept measurements across different target metals to anchor onset behavior. Implementing these efforts will improve the fidelity of performance measurements and extend hardware lifetime by managing SSE.

\section{Conclusions}

This work presents a comprehensive, atomistic investigation of neutral \ce{EMI-BF4} surface collisions on \ce{Au} extractor surfaces and implications for SSE and facility diagnostics in electrospray thrusters. Using energy-resolved QM/MM trajectories, we identify a three-stage sequence with increasing impact energy, ionic dissociation between $10$ to $20$ \unit{\eV}, neutralization between $30$ to $40$ \unit{\eV}, and covalent fragmentation greater than $50$ \unit{\eV}, and how to quantify how fragment composition, charge state, kinetic energy share, and scattering angle evolve across this window. The upper bound impacts at $100$ \unit{\eV} correspond to $0.5$ ${eV}/{u}$, falling right above the SEE onset and below the nominal SIE of $1$ ${eV}/{u}$ which help explain the emergence of charged secondaries in the ground testing \cite{geiger2025secondary}.

The trajectories confirm an operational point that neutral fragments pose danger to electrospray thrusters. In the intermediate energy regime, between $30$ to $40$ \unit{\eV} where neutralization pathways are dominant, large neutral populations with broad scattering angles are observed. Recent energy-swept \ce{Au} target experiments using a decelerating electrode directly corroborate this simulated result by showing charged-secondary emission when the ion plume is fully suppressed \cite{geiger2025secondary}. This demonstrates that neutral suppression alone does not eliminate SSE near the source, and, therefore, neutral management is needed.

The present QM/MM partition models the projectile and fragments quantum-mechanically and the \ce{Au} classically, meaning that explicit electron exchange with the metal and image-charge formation are not captured. Future work should focus on extending this model to include polarizable or QM metal layers, longer time windows for thermalization and desorption, variation of target surfaces and IL projectiles, and coupling to multiscale plume transport that will enable quantitative forecasts of beam divergence, extractor impacts, and facility bias with uncertainty bounds. Future experimental efforts should focus on energy-resolved tandem neutral and ion measurements that can anchor onset behavior and calibrate neutral diagnostics in order to advance the physical understanding and measurement fidelity required for flight-relevant performance predictions.

\section*{Acknowledgments}

The authors would like to graciously acknowledge the support of the Air Force Office of Scientific Research Young Investigator Program under Dr. Justin Koo, Grant No. FA9550-23-1-0141.

\section*{Author Contributions}

\textbf{Nicholas Laws}: Conceptualization (lead); Data curation (lead); Formal analysis (lead); Investigation (lead); Methodology (lead); Software (lead); Validation (lead); Visualization (lead); Writing – original draft (lead); Writing – review \& editing (lead).


\textbf{Elaine Petro}: Project administration (lead); Resources
(lead); Supervision (lead); Writing – review \& editing (supporting).

\nocite{*}
\bibliography{ref}

\end{document}